\documentclass[amsfonts,prd,aps]{revtex4}

\usepackage{graphicx}

\begin{document}

\title{New variables for 1+1 dimensional gravity}

\author{ Rodolfo Gambini$^{1}$,
Jorge Pullin$^{2}$,
Saeed Rastgoo$^{1}$}
\affiliation {
1. Instituto de F\'{\i}sica, Facultad de Ciencias,
Igu\'a 4225, esq. Mataojo, Montevideo, Uruguay. \\
2. Department of Physics and Astronomy, Louisiana State University,
Baton Rouge, LA 70803-4001}

\begin{abstract}
We show that the canonical formulation of a generic action for
$1+1$-dimensional models of gravity coupled to matter admits a
description in terms of Ashtekar-type variables. This includes the
CGHS model and spherically symmetric reductions of $3+1$ gravity as
particular cases. This opens the possibility of discussing models of
black hole evaporation using loop representation techniques and
verifying which paradigm emerges for the possible elimination of the
black hole singularity and the issue of information loss.
\end{abstract}

\maketitle
\section{Introduction}
Gravitational models in $1+1$ dimensions have proved a fertile ground
for testing ideas, in particular ideas about quantization. Examples of
these are the treatment of spherically symmetric reduction of $3+1$
models (see for example \cite{hato}) and also models intrinsic to
$1+1$ dimensions like the string-inspired
Callan--Giddings--Harvey--Strominger (CGHS) model of black hole
evaporation. This model has received renewed attention with the
construction of a new paradigm for its interpretation
\cite{asbo,astava}. These recent treatments however, have still been
done in terms of traditional quantization techniques. It would be
interesting to revisit them using the loop representation. Also
recently, we have made progress in treating spherically symmetric
reductions of general relativity using loop quantum gravity techniques
\cite{nosotros}. In particular we encounter that the singularity
inside black holes is eliminated and that the Fock vacuum emerges as a
quantum vacuum for a scalar field interacting in spherical symmetry
\cite{nosotros2}. The success of these techniques in the spherically
symmetric reduction of $3+1$ dimensional gravity strongly suggests
that such treatment should be extended to other $1+1$ dimensional
models, like the CGHS model, where the study of Hawking radiation and
black hole evaporation is tractable.  This allows to discuss the issue
of information loss and what paradigm describes better the final fate
of a quantum evaporating black hole.  In addition to this, $1+1$
dimensional models are the simplest models where one is faced in full
with the problem of dynamics of canonical quantum gravity and the
issue of the constraints forming an algebra with structure
functions. If one can deal with these problems, one may end up with a
full quantum gravitational description of an evaporating black hole,
one of the landmark problems of the field. With this goal, in this
paper we discuss how to canonically formulate these models using
Ashtekar-type variables. There have been treatments of spherically
symmetric reductions of $3+1$ gravity using such types of variables
\cite{bengtsson,thiemann,boka,bojowald,nosotros,nosotros2},
but we will here consider a fairly generic action that encompasses
many other $1+1$-dimensional models of interest.

We start with a quite general action in $1+1$ dimensions
\cite{verlinde}. We will follow the notation of \cite{strobl1},
\begin{equation}
S_{dil}=\int d^2x \sqrt{-|g|}\left(D(\Phi)R(g)
+\frac{1}{2}g^{ab}\partial_{a}\Phi \partial_{b}\Phi+U(\Phi)\right)
\label{eq:dil-full-g}.
\end{equation}
This is the most general diffeomorphism invariant action yielding
second order differential equations for the metric $g$ and a scalar
dilaton field $\Phi$. 

For further analysis, it is convenient to make a conformal transformation
$\tilde{g}_{ab}\equiv e^{\rho(\Phi)}g_{ab}$ with
\begin{equation}
\rho=\frac{1}{2}\int^{\Phi}\frac{du}{\frac{dD(u)}{du}}+\textrm{const.}\label{eq:rho-raw}
\end{equation}
followed by definition of a new field variable $X^3\equiv D(\Phi)$.

The superscript notation may appear strange but it is the case that it
is the third target space coordinate of a $\sigma$-model formulation
of the action and as such has been adopted in the literature
\cite{strobl1}, so we follow it here. With the above transformations
the action can be written as,
\begin{equation}
S_{\textrm{g-gen}}=\int d^{2}x \sqrt{-|\tilde{g}|}\left\{X^3 \tilde{R}+V(X^3)\right\}+S_{\textrm{m}},\label{eq:gen-action}
\end{equation}
where
\begin{equation}
V(z)\equiv\left(\frac{U}{\exp(\rho)}\right)(D^{-1}(z)),
\end{equation}
and this expression means $U/\exp(\rho)$ evaluated at $\Phi=D^{-1}(z)$.
$S_{\textrm{m}}$ represents the action of additional matter fields
one may wish to couple to the model.

There is a subtle issue that needs to be pointed out. In the following
calculations we will assume that $D$ has an inverse $D^{-1}$
everywhere on its domain of definition and, for simplicity, we assume
that $D$, $D^{-1}$, and $U$ are $C^{\infty}$. The final result,
however, will produce a set of Ashtekar-like variables for action
(\ref{eq:dil-full-g}) directly, independent of the invertibility or
not of $D$. This is important since both theories are generally
non-equivalent. In fact, 
with an action like (\ref{eq:gen-action}) it was not originally known how to
recover Hawking radiation in the model \cite{varadarajan,kurova}, 
although more careful analyses have shown how to recover it
\cite{vassilevich}. Previous treatments of the action 
(\ref{eq:gen-action}) with Ashtekar-type
variables have been considered by Bojowald and Reyes \cite{bore}.

Let us now illustrate two particular cases of interest. We start with
the spherically symmetric reduction of $3+1$ gravity in vacuum. We
choose as ansatz for the metric $ds^2=g_{ab}dx^{a}
dx^{b}+\Phi^2(d\theta^2+\sin^2(\theta)d\varphi^2)$, where $x^0$,
$x^1$, $\theta$ and $\varphi$ are coordinates adapted to the
spherical symmetry, $a,b=0,1$ and $g_{ab}(x^0,x^1)$ is the metric on
the $x^0,x^1$ plane.  Inserting this ansatz in the $3+1$ dimensional
Einstein--Hilbert action yields,
\begin{equation}
S_{\textrm{g-spher}}=\int d^2 x \sqrt{-|g|}\left(\frac{1}{4}\Phi^2 R(g)+\frac{1}{2}g^{ab}\partial_{a}\Phi \partial_{b}\Phi +\frac{1}{2} \right),\label{eq:g-spher}
\end{equation}
where $R$ is the Ricci tensor of the two dimensional metric $g^{ab}$ and $|g|$ is its determinant. In this case the choices are
\begin{equation}
D(\Phi)\equiv\frac{1}{4}\Phi^2,\quad
U(\Phi)\equiv\frac{1}{2}.\label{eq:D-U-spher}
\end{equation}
Using this and considering (\ref{eq:rho-raw}), we find for the conformal transformation
\begin{equation}
\rho=\ln(\Phi)\label{eq:rho-sph},
\end{equation}
and we can identify in the action that $V(X^3)=1/(4\sqrt{X^3})$.
We can then rewrite everything in
terms of $\Phi$:
\begin{eqnarray}
g_{ab}&=&\Phi^{-1}\tilde{g}_{ab}\\
\sqrt{-|g|}&=&\Phi^{-1}\sqrt{-|\tilde{g}|}\\
R&=&\Phi\tilde{R}+\tilde{g}^{ab}\partial_{a}\partial_{b}\Phi-
\Phi^{-1} \tilde{g}^{ab}\partial_{a}\Phi\partial_{b}\Phi.
\end{eqnarray}
One may choose
to minimally couple the model to a scalar field in $3+1$ dimensions, then one needs to add to the action a
term,
\begin{equation}
S_{\textrm{m-spher}}=-\int d^2x \sqrt{-|g|}\Phi^2 g^{ab}\partial_{a}f \partial_{b}f,
\end{equation}
where $f$ is the scalar field representing matter.

The other particular case of interest is the
Callan--Giddings--Harvey--Strominger (CGHS) model. It is given by the above
action with the choices
\begin{equation}
D(\Phi)\equiv\frac{1}{8}\Phi^2,\quad U(\Phi)\equiv\frac{1}{2}\Phi^2\lambda^{2}=4 D(\Phi)\lambda^{2}.\label{eq:D-U-CGHS}
\end{equation}
where $\lambda^2$ is a cosmological constant. The usual form of the CGHS action is obtained by introducing dilaton $\phi=-\ln(\Phi)/(2\sqrt{2})$,
yielding
\begin{equation}
S_{\textrm{g-CGHS}}=\int d^2x \sqrt{-|g|} e^{-2\phi}\left(R+4g^{ab}\partial_{a}\phi \partial_{b}\phi+4\lambda^{2}\right),\label{eq:g-CGHS}
\end{equation}
We introduce the conformal transformation.
Using (\ref{eq:D-U-CGHS}) and considering (\ref{eq:rho-raw}), we find
\begin{equation}
\rho=2\ln(\Phi)+c\label{eq:rho-CGHS}
\end{equation}

where $c$ is a constant. Choosing $c=-\ln(8)$ one reads off from the action that
$V(X^3)=4\lambda^2$.

The matter part of the action is,
\begin{equation}
S_{\textrm{m-CGHS}}=-\int d^2x \sqrt{-|g|}g^{ab}\partial_{a}f \partial_{b}f\equiv S_{\textrm{m-1+1}}\label{eq:m-CGHS}
\end{equation}
in which $f$ is the matter scalar field. It should be noted that this
form of the matter portion is not restricted to the CGHS model, but
corresponds to the coupling to a scalar field in $1+1$ dimensions no
matter what the model. It is only if one decides to couple a scalar field
in higher dimensions and then reduce that one gets a different action,
as we discussed above. To emphasize this point we will refer to it
from now on as $S_{\textrm{m-1+1}}$.

Let us introduce now a dyadic formulation for the gravitational part of the
generic action (\ref{eq:gen-action}),
\begin{equation}
L_g=-2X_I\epsilon^{ab}\left(\partial_{a}e_{b}{}^{I}+\omega_{a}\epsilon^{I}{}_{J}e_{b}{}^{J}\right) - 2X^3 \epsilon^{ab} \partial_{a}\omega_{b}+eV(X^3)\label{eq:L-g-gen}
\end{equation}
where $X_1$ and $X_2$ are Lagrange multipliers that make the theory
torsion-free. ``Internal'' indices $I,J$ range from $1,2$ and the
$e_a{}^I$ are dyads that we assume are invertible. $e$ is the
determinant of the dyad,
$e=\frac{1}{2}\epsilon^{ab}\epsilon_{IJ}e_{a}{}^{I}e_{b}{}^{J}$. $\epsilon_{IJ}$
and $\epsilon_{ab}$ are the two dimensional Levi--Civita symbols and
internal indices are raised an lowered with the $1+1$ dimensional
Minkowksi metric.  The matter part of the Lagrangian of
(\ref{eq:gen-action}) will read for spherically symmetric case:
\begin{equation}
L_{m\textrm{-spher}}=-4e X^3 \eta^{IJ}e^{a}{}_{I}e^{b}{}_{J}\partial_{a}f\partial_{b}f\label{eq:L-m-sph}
\end{equation}
and for the generic $1+1$ case,
\begin{equation}
L_{m\textrm{-1+1}}=-e\eta^{IJ}e^{a}{}_{I}e^{b}{}_{J}\partial_{a}f\partial_{b}f.\label{eq:L-m-CGHS}
\end{equation}
The Lagrangian (\ref{eq:L-g-gen}) can be rewritten, changing from space
to internal indices via,
\begin{eqnarray}
\epsilon^{ab}&=&-e \epsilon^{IJ} e^{a}{}_{I}e^{b}{}_{J},\\
\epsilon_{ab}&=&-e^{-1} \epsilon_{IJ} e_{a}{}^{I}e_{b}{}^{J},
\end{eqnarray}
as,
\begin{equation}
L_g=-2e\epsilon^{KI}e^{a}{}_{K}\partial_{a}X_{I}-2ee^{a I}X_{I} \omega_{a} + 2eX^{3}\epsilon^{KL} e^{a}{}_{K} e^{b}{}_{L} \partial_{a} \omega_{b}+eV(X^3)\label{eq:lag-bef-decomp}
\end{equation}

We now perform a $1+1$ decomposition of the action. We assume
space-time is sliced by surfaces $\Sigma_{t}$ parameterized by $t$.
Let $g_{ab}$ be the full space-time metric (strictly speaking in the
notation we introduced this should be $\tilde{g}_{ab}$, we omit the
tildes to simplify the notation) and $n_{a}$ be the unit timelike
vector field normal to the hypersurfaces $\Sigma_{t}$,
\begin{equation}
g_{ab}n^{a}n^{b}=-1.
\end{equation}
Then the space-time metric induces a spatial metric $q_{ab}$ on each $\Sigma_{t}$ such that
\begin{equation}
q_{ab}=g_{ab}+n_{a}n_{b},
\end{equation}
and this metric with one index raised $q^a_b=g^{ac}q_{cb}$ is a
projector on the spatial slice.  Introducing a vector field $t^a\equiv
(\partial/\partial t)^a$, we can define the shift vector $N^a=q^a{}_b
t^b$ and the lapse function $N=-g_{ab} t^a n^b$ such that
$n^a=(t^a-N^a)/N$.  One also has that $e=N\sqrt{q}$ where $q$ is the
determinant of the spatial part of the $1+1$ metric, i.e. $q=g_{11}$
The spatial components of $e_{a}{}^{I}$ can be written as
\begin{eqnarray}
E_{a}{}^{I}&=&q_{a}{}^{b}e_{b}{}^{I}\nonumber\\
&=&\left(g_{a}{}^{b}+n_{a}n^{b}\right)e_{b}{}^{I}\nonumber\\
&=&e_{a}{}^{I}+n_{a}n^{I}.
\end{eqnarray}
Substituting in the Lagrangian (\ref{eq:lag-bef-decomp}) we get,
\begin{eqnarray}
L_g=&-2N\sqrt{q}\Bigg\{&\left[E^{a}{}_{K}-\left(\frac{t^a - N^a}{N}\right)n_{K}\right]\partial_{a} {}^{*}X^K\nonumber\\
&&+ \left[E^{aI}-\left(\frac{t^a - N^a}{N}\right)n^{I}\right] \omega_{a}X_{I}\nonumber\\
&&- X^3 \epsilon^{KL} \left[E^{a}{}_{K}-\left(\frac{t^a - N^a}{N}\right)n_{K}\right] \left[E^{b}{}_{L}-\left(\frac{t^b - N^b}{N}\right)n_{L}\right] \partial_{a}\omega_{b}\nonumber\\
&&-\frac{V(X^3)}{2} \Bigg\}
\end{eqnarray}
where the dual of a tensor is defined by
${}^{*}X^K\equiv\epsilon^{KI}X_{I}$. The expression can  be reduced to
\begin{eqnarray}
L_g&=&-2N\tilde{E}^{a}{}_{K}\partial_{a} {}^{*}X^K + 2\sqrt{q} n_{K}t^{a}\partial_{a}{}^{*}X^K -2\sqrt{q} n_{K}N^{a}\partial_{a}{}^{*}X^K\nonumber\\
&&-2N\tilde{E}^{aI}X_{I}\omega_{a} + 2\sqrt{q} n^{I}X_{I}t^{a}\omega_{a} - 2\sqrt{q} n^{I}X_{I}N^{a}\omega_{a}\nonumber\\
&&-2X^3\epsilon^{KL}\tilde{E}^{a}{}_{K}t^{b}n_{L}\partial_{a}\omega_{b} - 2X^3\epsilon^{KL} \tilde{E}^{b}{}_{L} t^{a}n_{K} \partial_{a} \omega_{b} \nonumber\\
&&+N\sqrt{q}V(X^3).\label{eq:Lg-long-2}
\end{eqnarray}
where we have used
\begin{equation}
\sqrt{q}E^{a}{}_{I}=\tilde{E}^{a}{}_{I}.
\end{equation}
Noticing
\begin{eqnarray}
\mathcal{L}_{t}\omega_{b}&=&t^{a}\partial_{a}\omega_{b}+\omega_{a}\partial_{b}t^{a}\\
\mathcal{L}_{t}{}^{*}X^K&=&t^{a}\partial_{a}{}^{*}X^K,
\end{eqnarray}
the gravitational Lagrangian becomes
\begin{eqnarray}
L_g&=&-2N\tilde{E}^{a}{}_{K}\partial_{a} {}^{*}X^K + 2\sqrt{q} n_{K} \mathcal{L}_{t}{}^{*}X^K -2\sqrt{q} n_{K} N^{a} \partial_{a}{}^{*}X^K \nonumber\\
&&-2N\tilde{E}^{aI}X_{I}\omega_{a} + 2\sqrt{q} n^{I}X_{I}t^{a}\omega_{a} - 2 \sqrt{q} n^{I}X_{I} N^{a} \omega_{a}\nonumber\\
&&- 2X^3 \epsilon^{KL}\tilde{E}^{b}{}_{L}n_{K} \left(\mathcal{L}_{t}\omega_{b}-\partial_{b} (t^{a}\omega_{a})\right)\nonumber\\
&&+N\sqrt{q}V(X^3).\label{eq:Lg-clean1}
\end{eqnarray}
which can be rewritten as,
\begin{eqnarray}
L_g&=&-2N {}^{*}n_{I} D_{1} {}^{*}X^I - 2\sqrt{q} n_{I} N^{1} D_{1}{}^{*}X^I + 2\sqrt{q} n_{I} {}^{*}\dot{X}^I \nonumber\\
&& + \omega_{0}\big[2\sqrt{q} n_{I}X^{I}+\partial_{1} \big(2X^3 \big)\big] + 2X^3  \left(\dot{\omega}_{1}\right)\nonumber\\
&&+N\sqrt{q}V(X^3),\label{eq:Lg-clean3}
\end{eqnarray}
where the derivative operator $D_{a}$ is defined by
$D_{a}X_{I}=\partial_{a}X_{I}+\omega_{a}\epsilon_{I}{}^{J}X_{J}$.  We
have chosen adapted coordinates $x^0=t$, $x^1=x$ with the standard
basis vectors, so $t^0=1$ and $t^1=0$ and therefore the Lie derivative
becomes an ordinary derivative which we denotes by a dot.  To derive
the above expression we expand $\tilde{E}^{1}{}_{I}$ in terms of some
orthonormal vector field in the tangent space of the spatial
hypersurface. One can see that the dual of $n^{I}$ vector field can be
a candidate:
\begin{eqnarray}
{}^{*}n_{I}{}^{*}n^{I}&=&\epsilon_{IK} n^{K} \epsilon^{IL}n_{L}\nonumber\\
&=&-\delta_{K}{}^{L} n^{K} n_{L}\nonumber\\
&=&-n^{K} n_{K}\nonumber\\
&=&1.
\end{eqnarray}
So we are able to expand $\tilde{E}^{1}{}_{I}$ as
\begin{equation}
\tilde{E}^{1}{}_{I}=\tilde{E}^{1}{}_{\parallel}{}^{*}n_{I},\label{eq:E-nstar}
\end{equation}
in which $\tilde{E}^{1}{}_{\parallel}$ is the only component of $\tilde{E}^{1}{}_{I}$ with respect to the basis vector field ${}^{*}n_{I}$. One can also see that for the $\tilde{E}^{1}{}_{I}$ field:
\begin{eqnarray}
1&=&\eta^{IJ}\tilde{E}^{1}{}_{I}\tilde{E}^{1}{}_{J}\nonumber\\
&=&\eta^{IJ}\tilde{E}^{1}{}_{\parallel}{}^{*}n_{I} \tilde{E}^{1}{}_{\parallel}{}^{*}n_{J}\nonumber\\
&=&\big(\tilde{E}^{1}{}_{\parallel}\big)^{2},
\end{eqnarray}
and thus
\begin{equation}
\tilde{E}^{1}{}_{\parallel}=1.\label{eq:Ep-eq1}
\end{equation}
Using (\ref{eq:E-nstar}) and (\ref{eq:Ep-eq1}), we can rewrite $\tilde{E}^{1}{}_{I}$ as
\begin{eqnarray}
\tilde{E}^{1}{}_{I}&=& \tilde{E}^{1}{}_{\parallel}{}^{*}n_{I}\nonumber\\
&=&{}^{*}n_{I}.\label{eq:Etilde-q-1}
\end{eqnarray}
Based on this we can arrive at a useful observation:
\begin{eqnarray}
\epsilon^{KL}\tilde{E}^{1}{}_{L}n_{K}&=&\epsilon^{KL} \tilde{E}^{1}{}_{\parallel}{}^{*}n_{L}n_{K}\nonumber\\
&=&\epsilon^{KL} {}^{*}n_{L}n_{K}\nonumber\\
&=& \epsilon^{KL} \epsilon_{LJ} n^{J}n_{K}\nonumber\\
&=&- \epsilon^{LK} \epsilon_{LJ} n^{J}n_{K}\nonumber\\
&=&- (-\delta_{J}{}^{K}) n^{J}n_{K}\nonumber\\
&=& n^{K}n_{K}\nonumber\\
&=&-1.\label{eq:Etilde-q-2}
\end{eqnarray}
and using this we get
\begin{eqnarray}
2X^3 \epsilon^{KL}\tilde{E}^{1}{}_{L}n_{K}\partial_{1} \omega_{0}&=&\partial_{1}\big(2X^3 \epsilon^{KL}\tilde{E}^{1}{}_{L}n_{K} \omega_{0} \big) - \omega_{0} \partial_{1} \big(2X^3 \epsilon^{KL}\tilde{E}^{1}{}_{L}n_{K}\big) \nonumber\\
&=&-\partial_{1}\big(2X^3 \omega_{0} \big) + \omega_{0} \partial_{1} \big(2X^3 \big). \label{eq:Etilde-omega0}
\end{eqnarray}

Going back to the Lagrangian (\ref{eq:Lg-clean3}) one can immediately
identify the canonical variables ${}^{*}X^I$, $I=1,2$ and $\omega_1$,
and their canonical momenta,
\begin{eqnarray}
P_{I}&=&\frac{\partial L_g}{\partial\, {}^{*}\dot{X}^I} = 2\sqrt{q} n_{I},\label{eq:eqmot-Lg-PI}\\
P_{3}&=&\frac{\partial L_g}{\partial \dot{\omega}_{1}} = 2X^3.\label{eq:eqmot-Lg-P3}
\end{eqnarray}
One can then rewrite the Lagrangian in terms of the canonical variables,
\begin{eqnarray}
L_g&=&-2N \epsilon_{IJ} \frac{P^{J}}{\|P\|} D_{1} {}^{*}X^I - P_{I} N^{1} D_{1}{}^{*}X^I + P_{I} {}^{*}\dot{X}^I \nonumber\\
&& + \omega_{0}\big[P_{I} \epsilon^{IJ} {}^{*}X_{J}+\partial_{1} \big(P_{3} \big)\big] + P_{3}  \left(\dot{\omega}_{1}\right)\nonumber\\
&&+N\frac{\|P\|}{2}V(X^3).\label{eq:Lg-clean-fin}
\end{eqnarray}
where
\begin{equation}
\|P\|^2=-\eta^{IJ}P_{I}P_{J}=-4 q \eta^{IJ} n_{I} n_{J}=4q.
\end{equation}
{}For the gravitational part one can therefore write the total Hamiltonian in the generic case,
\begin{eqnarray}
H_{\textrm{gen}}&=&N\left(2 \epsilon_{IJ} \frac{P^{J}}{\|P\|} D_{1} {}^{*}X^I - \frac{\|P\|}{2}V(P^3) \right)\nonumber\\
&&+N^{1} \left( P_{I} D_{1}{}^{*}X^I \right) - \omega_{0}\left(P_{I} \epsilon^{IJ} {}^{*}X_{J}+\partial_{1} \big(P_3 \big)\right).
\end{eqnarray}

Let us now turn to the matter Lagrangians. Denoting $\partial_1
f\equiv f'$ and $\partial_t f\equiv \dot{f}$ the matter part of the
Lagrangian for the spherical symmetric reduction of $3+1$ gravity can
be written as,
\begin{equation}
L_{m\textrm{-spher}}=4\sqrt{q} X^3 \bigg\{-\frac{N}{q}f'^{2} + \frac{1}{N}\dot{f}^{2} -\frac{2}{N} N^{1}\dot{f}f' +\frac{1}{N} (N^{1})^{2}f'^{2} \bigg\}.\label{eq:pickhere}
\end{equation}
Now it is easy to read the canonical variable $f$ and its conjugate being
\begin{equation}
P_{f}=\frac{\partial L_{m\textrm{-spher}}}{\partial \dot{f}}=\frac{8\sqrt{q} X^3 }{N}\big[\dot{f}-N^{1}f' \big].\label{eq:Pf}
\end{equation}
which leads to
\begin{equation}
H_{m\textrm{-spher}}=\frac{4 N P_3 f'^{2}}{\|P\|} + \frac{N(P_{f})^{2}}{4\|P\|P_3} + N^1 f' P_{f}.\label{eq:spher-matter-fin}
\end{equation}

For the generic $1+1$ case, the matter Lagrangian differs from the one
we just considered in a factor $4X^3$, so very straightforwardly one
gets,
\begin{equation}
H_{m\textrm{-1+1}}=\frac{2N f'^{2}}{\|P\|} + \frac{N(P_{f})^{2}}{2\|P\|} + N^1 f' P_{f} \label{eq:CGHS-matter-fin}
\end{equation}

The total Hamiltonian including matter for the generic $1+1$ case is,
\begin{eqnarray}
H_{\textrm{gen}}&=&N\left(2 \epsilon_{IJ} \frac{P^{J}}{\|P\|} D_{1} {}^{*}X^I - \frac{\|P\|}{2}V(P^3) +\frac{2 f'^{2}}{\|P\|} + \frac{(P_{f})^{2}}{2\|P\|}\right)\nonumber\\
&&+N^{1} \left( P_{I} D_{1}{}^{*}X^I + f' P_{f} \right) - \omega_{0}\left(P_{I} \epsilon^{IJ} {}^{*}X_{J}+\partial_{1} \big(P_3 \big)\right).
\end{eqnarray}
The generic analysis can be carried out only up to this point since
the conformal transformation leading from the original metric
variables to those with tildes given by equation (\ref{eq:rho-raw})
involves an arbitrary function $D(\Phi)$.  To complete the
construction of the Ashtekar-type variables one needs to specify such
function.

Let us now introduce Ashtekar-like variables for both cases, starting
with the spherical reduction of $3+1$ gravity.
We first notice that in this case $E^x=\Phi^2=4X^3=4D(\Phi)$ with
$E^x$ the densitized triad in the radial direction, as can be readily seen
from the form of the spherically symmetric metric. On the other hand
${q}=\tilde{g}_{11}=(E^\varphi)^2/\sqrt{E^x}$.
The components $1,2$ of the normal vector $n_I$ form a vector in
the ``transverse'' space to the radial direction that is normalized so they
can be parameterized by an angle $\eta$,
\begin{eqnarray}
n_{1}&=&\cosh(\eta)\label{eq:asht-canon-n1}\\
n_{2}&=&\sinh(\eta)\label{eq:asht-canon-n2}.
\end{eqnarray}
We can now introduce the densitized triad in the $\varphi$ direction
by using its relation to the determinant of the three metric,
\begin{equation}
\sqrt{q}=\frac{E^{\varphi}}{(E^{x})^{\frac{1}{4}}}=\frac{\|P\|}{2}\label{eq:asht-q-ex-pnorm}.
\end{equation}
We also note from (\ref{eq:eqmot-Lg-PI}) and (\ref{eq:asht-q-ex-pnorm}) that
\begin{eqnarray}
P_{1}&=&n_1 \|P\|=\frac{2E^{\varphi}}{(E^{x})^{\frac{1}{4}}}\cosh(\eta)\label{eq:asht-canon-P1}\\
P_{2}&=&n_2 \|P\|=\frac{2E^{\varphi}}{(E^{x})^{\frac{1}{4}}}\sinh(\eta)\label{eq:asht-canon-P2}\\
P_3&=&\frac{E^{x}}{2}\label{eq:P3-Ex}.
\end{eqnarray}
and with the above relations we can motivate a type II canonical transformation
that will leave us with canonical variables $E^x,E^\varphi, \eta$ and
their canonically conjugates  which we will call $A_x,K_\varphi, Q_\eta$.
The generating function is
\begin{equation}
F(q,P)={}^{*}X^{1}\frac{2E^{\varphi}}{(E^{x})^{\frac{1}{4}}}\cosh(\eta)+{}^{*}X^{2}\frac{2E^{\varphi}}{(E^{x})^{\frac{1}{4}}}\sinh(\eta)+\omega_{1} \frac{E^{x}}{2},
\end{equation}
and it leads to the following expressions for the new canonical variables,
\begin{eqnarray}
Q_{\eta}&=&\frac{\partial F}{\partial \eta}=\frac{2E^{\varphi}}{(E^{x})^{\frac{1}{4}}}({}^{*}X^1\sinh(\eta)+{}^{*}X^2\cosh(\eta))=-{}^{*}X_1 P_2 + {}^{*}X_2 P_1\\
K_{\varphi}&=&\frac{\partial F}{\partial E^{\varphi}}=\frac{2{}^{*}X^1\cosh(\eta)+2{}^{*}X^2\sinh(\eta)}{(E^{x})^{\frac{1}{4}}}=\frac{-{}^{*}X_1 P_1 + {}^{*}X_2 P_2}{E^{\varphi}}\\
A_x&=&\frac{\partial F}{\partial E^{x}} =-\frac{E^{\varphi}}{2(E^{x})^{\frac{5}{4}}}({}^{*}X^1\cosh(\eta)+{}^{*}X^2\sinh(\eta))+\frac{\omega_1}{2}\\
&=&\frac{{}^{*}X_1 P_1 - {}^{*}X_2 P_2}{4E^{x}}+\frac{\omega_1}{2}\\
&=&-\frac{E^{\varphi}K_{\varphi}}{4E^{x}}+\frac{\omega_1}{2}
\end{eqnarray}
where we have used ${}^{*}X^{1}=-{}^{*}X_{1}$ and ${}^{*}X^{2}={}^{*}X_{2}$. We still need to find expressions for ${}^{*}X_{1}$ and ${}^{*}X_{2}$ in terms of these new variables. One can see that,
\begin{eqnarray}
{}^{*}X_{1}&=&\frac{Q_{\eta}(E^{x})^{\frac{1}{4}}\sinh(\eta)}{2E^{\varphi}}-\frac{K_{\varphi}(E^{x})^{\frac{1}{4}}\cosh(\eta)}{2}\label{eq:X1star},\\
{}^{*}X_{2}&=&\frac{Q_{\eta}(E^{x})^{\frac{1}{4}}\cosh(\eta)}{2E^{\varphi}}-\frac{K_{\varphi}(E^{x})^{\frac{1}{4}}\sinh(\eta)}{2}\label{eq:X2star}.
\end{eqnarray}
We are now ready to express the total Hamiltonian in terms of the new variables,
\begin{eqnarray}
H_{\textrm{spher}}&=&N\bigg[\frac{1}{2}(E^{x}(x))^{\frac{1}{4}}E^{\varphi}(x)(f'(x))^2 + \frac{1}{4}\frac{Q_{\eta}(x){E^{x}}'(x)}{(E^{x}(x))^{\frac{3}{4}}E^{\varphi}(x)} - \frac{(E^{x}(x))^{\frac{1}{4}} Q_{\eta}(x) {E^{\varphi}}'(x)} {(E^{\varphi}(x))^2} \nonumber\\
&& +\frac{(E^{x}(x))^{\frac{1}{4}}Q'_{\eta}(x)}{E^{\varphi}(x)} - (E^{x}(x))^{\frac{1}{4}}K_{\varphi}(x)\eta'(x) - 2 (E^{x}(x))^{\frac{1}{4}}A_x(x) K_{\varphi}(x) \nonumber\\
&&- \frac{1}{2}\frac{E^{\varphi}(x)(K_{\varphi}(x))^2}{(E^{x}(x))^{\frac{3}{4}}} - \frac{1}{2} \frac{E^{\varphi}(x)}{(E^{x}(x))^{\frac{3}{4}}} + \frac{1}{4} \frac{(P_f(x))^2}{E^{\varphi}(x)(E^{x}(x))^{\frac{1}{4}}} + \frac{(E^{x}(x))^{\frac{5}{4}}(f'(x))^2}{E^{\varphi}(x)}\bigg]\nonumber\\
&&+\omega_{0}\bigg[Q_{\eta}(x)-\frac{1}{2}{E^{x}}'(x)\bigg]\nonumber\\
&&+N^1\bigg[-Q_{\eta}(x)\eta'(x) + \frac{1}{4} \frac{K_{\varphi}(x){E^{x}}'(x)E^{\varphi}(x)}{E^{x}(x)}+K'_{\varphi}(x) E^{\varphi}(x) - 2A_x(x)Q_{\eta}(x) \nonumber\\
&&-\frac{1}{2}\frac{E^{\varphi}(x)K_{\varphi}(x)Q_{\eta}(x)}{E^{x}(x)} + P_f(x)f'(x)\bigg]
\end{eqnarray}
and we readily distinguish a Hamiltonian and diffeomorphism constraints
and a Gauss law.  One can proceed further by solving Gauss' law
$Q_{\eta}(x)=\frac{1}{2}{E^{x}}'(x)$ and defining a new variable
$K_x(x)=\frac{1}{2}\eta'(x)+A_x(x)$, where one is left with a model with a
Hamiltonian and diffeomorphism constraint and with canonical pairs
$E^x,K_x$ and $E^\varphi,K_\varphi$. We will not repeat the calculation
here since it is already present in the literature \cite{bojowald,nosotros}.

Let us now turn our attention to the CGHS model. The construction is virtually similar, except for small details. We first introduce $E^x$ and the angle $\eta$,
\begin{eqnarray}
X^{3}&=&D(\Phi)=\frac{E^{x}}{8}\label{eq:asht-x3-ex-CGHS}\\
\sqrt{q}&=&E^{\varphi}=\frac{\|P\|}{2}\label{eq:asht-q-ex-pnorm-CGHS}\\
n_{1}&=&\cosh(\eta)\label{eq:asht-canon-n1-CGHS}\\
n_{2}&=&\cosh(\eta)\label{eq:asht-canon-n2-CGHS}
\end{eqnarray}
and
\begin{equation}
\|P\|=2E^{\varphi}\label{eq:normP-CGHS}
\end{equation}
and from (\ref{eq:eqmot-Lg-PI}) and (\ref{eq:asht-q-ex-pnorm-CGHS})
\begin{eqnarray}
P_{1}&=&n_1 \|P\|=2E^{\varphi}\cosh(\eta)\label{eq:asht-canon-P1-CGHS},\\
P_{2}&=&n_2 \|P\|=2E^{\varphi}\sinh(\eta)\label{eq:asht-canon-P2-CGHS},\\
P_3&=&\frac{E^{x}}{4}\label{eq:P3-Ex-CGHS}
\end{eqnarray}
and the generating function for the type II canonical transformation is,
\begin{equation}
F(q,P)={}^{*}X^{1}2E^{\varphi}\cosh(\eta)+{}^{*}X^{2}2E^{\varphi}\sinh(\eta)+\omega_{1} \frac{E^{x}}{4},
\end{equation}
The new canonical variables are then
\begin{eqnarray}
Q_{\eta}&=&\frac{\partial F}{\partial \eta}=2E^{\varphi}({}^{*}X^1\sinh(\eta)+{}^{*}X^2\cosh(\eta))=-{}^{*}X_1 P_2 + {}^{*}X_2 P_1\\
K_{\varphi}&=&\frac{\partial F}{\partial E^{\varphi}}=2{}^{*}X^1\cosh(\eta)+2{}^{*}X^2\sinh(\eta)=\frac{-{}^{*}X_1 P_1 + {}^{*}X_2 P_2}{E^{\varphi}}\\
A_x&=&\frac{\partial F}{\partial E^{x}} =\frac{\omega_1}{4}
\end{eqnarray}
and the total Hamiltonian is,
\begin{eqnarray}
H_{\textrm{CGHS}}&=&N\bigg[\frac{(f'(x))^2}{E^{\varphi}(x)} - \frac{Q_{\eta}(x) {E^{\varphi}}'(x)} {(E^{\varphi}(x))^2} +\frac{Q'_{\eta}(x)}{E^{\varphi}(x)} - K_{\varphi}(x)\eta'(x)\nonumber\\
&& - 4A_x(x) K_{\varphi}(x) - 4 E^{\varphi}(x)\lambda^{2} + \frac{1}{4} \frac{(P_f(x))^2}{E^{\varphi}(x)}\bigg]\nonumber\\
&&+\omega_{0}\bigg[Q_{\eta}(x)-\frac{1}{4}{E^{x}}'(x)\bigg]\nonumber\\
&&+N^1\bigg[-Q_{\eta}(x)\eta'(x) +K'_{\varphi}(x) E^{\varphi}(x) - 4A_x(x)Q_{\eta}(x) + P_f(x)f'(x)\bigg]
\end{eqnarray}
where again we easily distinguish the Gauss law, Hamiltonian and
diffeomorphism constraint. We can gauge fix Gauss' law
$Q_{\eta}(x)=\frac{1}{4}{E^{x}}'(x)$ and defining
$K_x(x)=\frac{1}{4}\eta'(x)+A_x(x)$ we are left with a total Hamiltonian,
\begin{eqnarray}
H_{\textrm{CGHS}}&=&N\mathcal{H}+N^{1}\mathcal{D}\nonumber\\
&=&N\bigg[-\frac{1}{4}\frac{{E^{x}}'(x){E^{\varphi}}'(x)}{(E^{\varphi}(x))^{2}} + \frac{1}{4}\frac{{E^{x}}''(x)}{E^{\varphi}(x)}- 4 K_{\varphi}(x) K_{x}(x) -4 E^{\varphi}(x)\lambda^{2} \nonumber\\
&&+\frac{(f'(x))^{2}}{E^{\varphi}(x)} +\frac{1}{4}\frac{(P_{f}(x))^{2}}{E^{\varphi}(x)}\bigg]\nonumber\\
&&+N^{1}\bigg[-{E^{x}}'(x)K_x(x)+E^{\varphi}(x)K'_{\varphi}(x)+P_f(x)f'(x)\bigg]\label{eq:Hamilt-final-CGHS}
\end{eqnarray}

We therefore have developed a technique for constructing Hamiltonian
formulation for generic models of gravity in $1+1$ dimensions in terms
of Ashtekar variables. 
This includes the CGHS black hole model.  These
will be the point of departure for future investigations that will
probe the emergence of a paradigm in which black hole singularities
are eliminated and the information loss problem in evaporating black
holes can be treated using the loop representation. The techniques are
ready. The following steps would be to polymerize the above
expressions and a study of the resulting semiclassical theory to study
modifications or perhaps the elimination of the singularity. This can
be followed by studies in the full quantum theory using the uniform
discretization procedure, as has been carried out in the spherical
case already \cite{nosotros2}, where the vacuum has been identified
and progress is being made on computing propagators.

We wish to thank Dmitri Vassilevich for comments on a previous version
of the manuscript. This work was supported in part by grant
NSF-PHY-0650715, funds of the Hearne Institute for Theoretical
Physics, FQXi, CCT-LSU, Pedeciba and ANII PDT63/076.


\begin{thebibliography}{99}
\bibitem{hato}
  P.~Thomi, B.~Isaak and P.~Hajicek,
  Phys.\ Rev.\  D {\bf 30}, 1168 (1984).
\bibitem{asbo}
  A.~Ashtekar and M.~Bojowald,
  Class.\ Quant.\ Grav.\  {\bf 22}, 3349 (2005)
  [arXiv:gr-qc/0504029].
\bibitem{astava}
  A.~Ashtekar, V.~Taveras and M.~Varadarajan,
  Phys.\ Rev.\ Lett.\  {\bf 100}, 211302 (2008)
  [arXiv:0801.1811 [Unknown]].
\bibitem{nosotros}
  M.~Campiglia, R.~Gambini and J.~Pullin,
  Class.\ Quant.\ Grav.\  {\bf 24}, 3649 (2007)
  [arXiv:gr-qc/0703135].

\bibitem{nosotros2}
  R.~Gambini, J.~Pullin and S.~Rastgoo,
  arXiv:0906.1774 [gr-qc].
\bibitem{bengtsson}
 I.~Bengtsson,
  Class.\ Quant.\ Grav.\  {\bf 5}, L139 (1988).

\bibitem{thiemann}
T. Thiemann, H. Kastrup, Nucl. Phys. {\bf B399}, 211 (1993);
\bibitem{boka}
  M.~Bojowald and H.~A.~Kastrup,
  Class.\ Quant.\ Grav.\  {\bf  17}, 3009 (2000)
  [arXiv:hep-th/9907042].


\bibitem{bojowald}
  M.~Bojowald and R.~Swiderski,
  Class.\ Quant.\ Grav.\  {\bf 23}, 2129 (2006)
  [arXiv:gr-qc/0511108].



\bibitem{verlinde} H. Verlinde, in
``6th Marcel Grossmann Meeting on General Relativity'', M. Sato, editor, World Scientific,
Singapore (1992).
\bibitem{strobl1}
T.~Klosch and T.~Strobl,
  Class.\ Quant.\ Grav.\  {\bf 13}, 965 (1996)
  [Erratum-ibid.\  {\bf 14}, 825 (1997)]
  [arXiv:gr-qc/9508020].

\bibitem{varadarajan}
  M.~Varadarajan,
  Phys.\ Rev.\  D {\bf 57}, 3463 (1998)
  [arXiv:gr-qc/9801058].
\bibitem{kurova}
  K.~V.~Kuchar, J.~D.~Romano and M.~Varadarajan,
  Phys.\ Rev.\  D {\bf 55}, 795 (1997)
  [arXiv:gr-qc/9608011].

\bibitem{vassilevich}
  W.~Kummer and D.~V.~Vassilevich,
  Annalen Phys.\  {\bf 8}, 801 (1999)
  [arXiv:gr-qc/9907041]

  D.~Grumiller, W.~Kummer and D.~V.~Vassilevich,
  Phys.\ Rept.\  {\bf 369}, 327 (2002)
  [arXiv:hep-th/0204253].

\bibitem{bore}
  M.~Bojowald and J.~D.~Reyes,
  Class.\ Quant.\ Grav.\  {\bf 26}, 035018 (2009)
  [arXiv:0810.5119 [gr-qc]].
\end{thebibliography}
\end{document}